\documentclass[10pt]{iopart}
\usepackage{iopams} 
\usepackage{setstack}
\usepackage{graphics}
\usepackage{amsthm, graphicx}
\usepackage[colorlinks=true,linkcolor=blue,urlcolor=blue,citecolor=blue]{hyperref}

\usepackage{bm}
\usepackage{amsthm}
\usepackage{hyperref}
\usepackage{xcolor}
\hypersetup{
    urlcolor=blue,
    citecolor=blue
           }

\theoremstyle{plain}
\usepackage{color}
\usepackage{amssymb}
\usepackage{amsthm}
\usepackage{float}
\usepackage{tabularx}
\usepackage{graphicx}
\usepackage[utf8]{inputenc}

\usepackage{esvect}
\usepackage{wrapfig}
\usepackage{amsthm}
\usepackage{verbatim}
\usepackage{bbm}
\usepackage[normalem]{ulem}

\usepackage{enumitem}
\usepackage{fmtcount}
\usepackage{booktabs}

\usepackage{csquotes}
\usepackage{epsfig}
\usepackage{tabularx}
\usepackage{graphicx}

\usepackage{latexsym}
\usepackage{bm}
\usepackage{graphics,epstopdf}
\usepackage{enumitem}
\usepackage{fmtcount}
\usepackage{booktabs}
\usepackage{csquotes}
\usepackage{epsfig}
\usepackage[normalem]{ulem}
\usepackage{float}
\usepackage{graphicx}  
\usepackage{dcolumn}          
\usepackage{amssymb}
\usepackage{appendix}
\usepackage{esvect}
\usepackage{wrapfig}
\usepackage{amsthm}
\usepackage{verbatim}
\usepackage{bbm}
\usepackage[mathscr]{euscript}

\newtheorem{theorem}{Theorem}
\newtheorem*{theorem*}{Theorem}
\newtheorem*{question*}{Question}

\newcommand{\bra}[1]{\ensuremath{\left\langle #1\right|}}
\newcommand{\ket}[1]{\ensuremath{\left|#1\right\rangle}}
\newcommand{\braket}[2]{\ensuremath{\left\langle #1\vphantom{#2}\right.\left|\vphantom{#1}#2\right\rangle}}
\newcommand{\ketbra}[2]{\ensuremath{\left| #1\vphantom{#2}\right\rangle\left\langle\vphantom{#1}#2\right|}}

\begin{document}
\title[Role of energy-invariant assistants in energy extraction from quantum batteries]{Role of energy-invariant {assistants} in energy extraction from quantum batteries}

\author{Paranjoy Chaki}
\address{Harish-Chandra Research Institute,  A CI of Homi Bhabha National Institute, Chhatnag Road, Jhunsi, Allahabad 211 019, India}

\author{Aparajita Bhattacharyya}
\address{Harish-Chandra Research Institute,  A CI of Homi Bhabha National Institute, Chhatnag Road, Jhunsi, Allahabad 211 019, India}

\author{Kornikar Sen\footnote{Corresponding author}}
\address{Harish-Chandra Research Institute,  A CI of Homi Bhabha National Institute, Chhatnag Road, Jhunsi, Allahabad 211 019, India}
\address{Departamento de Física Teórica, Universidad Complutense, 28040 Madrid, Spain}
\ead{kornikar.sen@gmail.com}

\author{Ujjwal Sen}
\address{Harish-Chandra Research Institute,  A CI of Homi Bhabha National Institute, Chhatnag Road, Jhunsi, Allahabad 211 019, India}

\begin{abstract}
We investigate the role of {energy-invariant assistants} in energy extraction from quantum batteries. To this end, for energy extraction, we restrict to unitaries that jointly act on the battery and the {assistant} but keep the {assistant’s} energy preserved. We demonstrate that in the presence of an energy-invariant assistant, having the same dimension as the battery, all stored energy of the battery can always be extracted, transforming the battery into its ground state, when an appropriate joint unitary and assistant's state are employed. Additionally, we offer a necessary and sufficient condition for a battery to be unable to provide any energy, i.e., to be inactive, even when an energy-invariant assistant is present, prepared in an arbitrary but fixed state.   
\end{abstract}
\maketitle

\section{Introduction}\label{sec1}
The demand for small-sized electrical gadgets has increased immensely in the modern era due to advances in nanotechnology and the utilization of small-scale electronic devices.  
Miniaturization of devices enhances quantum mechanical features in them, which has led scientists to consider quantum versions of technological devices such as refrigerators~\cite{refregirator,Sk_ref,Das_2019}, engines~\cite{engine,Eng_3}, batteries~\cite{Alicki_2013,Alhambra,npovm}, etc.   
A battery that stores electrical energy that can be extracted whenever needed is one of the most widely used electrical instruments. This specific instrument finds extensive usage in workplaces, industries, transportation, etc.  

The quantum mechanical prototype of a battery was first provided by R. Alicki and M. Fannes~\cite{Alicki_2013} along with the concept of ergotropy~\cite{Ergotropy,batt_rev2,capacity2}. The ergotropy of a quantum battery is the maximum possible extractable energy from the battery by the action of unitaries on it. At the same time, the concept of daemonic ergotropy is given in~\cite{ord_demon}. {Ergotropy's relation with the concept of passive states is first provided in Ref.~\cite{Pusz_1978}. In Ref.~\cite{passive3}, authors identify the passive states that lie at the opposite extreme—those that maximize energy for a given entropy and, equivalently, minimize entropy for a fixed energy. In Ref.~\cite{passive4}, authors provide a proof showing that only Gibbs states at positive temperatures are completely passive and introduce the notion of virtual temperatures, associated with energy-level pairs, to show that passive states correspond to transitions with positive temperatures, while complete passivity arises when all transitions share an identical positive temperature. In Ref.~\cite{passive5}, the authors examine the limitations of Gaussian unitaries for work extraction and introduce the concept of Gaussian passivity, providing necessary and sufficient conditions for states whose energy cannot be reduced by Gaussian operations. The necessary and sufficient condition of passivity under local CPTP operation is given in \cite{Passive8}. On the other hand, in Ref.~\cite{passive6}, the authors consider a single system and show that all passive states, except thermal ones, are unstable under weaker reversible operations. From any athermal state, optimal energy can be extracted using a reversible cycle, implying that only completely passive (thermal) states remain stable under general reversible processes outside the thermodynamic limit. The necessary and sufficient condition of local passivity was provided in Ref.~\cite{ref20}.}

Extraction of the entire accessible energy from a system using unitaries on it transforms the system to its passive state, i.e., a state from which no further energy extraction is possible using unitary operations. Apart from that, energy extraction through local non-completely positive maps have also been investigated~\cite{ncptp}.

In addition to energy extraction, there has been significant research on the charging of quantum such Ref.~\cite{charg_m} deals with $N$ qubits and demonstrates that the inclusion of global operations enables an N-fold increase in the power generated per qubit. In Ref.~\cite{charging_3}, authors show that the presence of dephasing noise causes fast charging of quantum batteries.  
A quantum battery can be charged by applying a unitary on the battery, that is through switching on a field or interaction between different parts of the battery~\cite{bat_rev1}. 
 Another efficient technique for charging a quantum battery is coupling an auxiliary system, often referred to as a charger, with it and transferring energy from the auxiliary to the battery through an interaction~\cite{charger_1st,auxi,Charging_4}. 

Various many-body systems, including the short- and long-range XXZ quantum spin model~\cite{xyz2}, the Hubbard model, bosonic and fermionic systems~\cite{PhysRevA.106.022618}, spin cavity models~\cite{cavitya}, non-Hermitian systems~\cite{non-hermitian}, etc., have been analyzed to study the charging and discharging capabilities of different quantum battery models. In addition, a comparison of many-body quantum and classical systems has been carried out~\cite{classical}. The role of ordered and disordered interaction in the performance of a quantum battery is observed in Ref.~\cite{h_b}. {On the other hand a no-go theorem in the context of the universal charging protocol of quantum batteries is given in Ref.~\cite{univ_c}.} Apart from that, the significance of the dimension of quantum batteries in charging them in the presence of impurities has been explored~\cite{Dim_E}. Fast charging of quantum batteries assisted by noise has been observed in Ref.~\cite{Fq}.
Simultaneously, there are works which have addressed the role of quantum correlations like entanglement~\cite{Entt_1,ent2,demon1} and coherence~\cite{coh_1,coh_2,coh_3} in the performance of quantum batteries. In order to investigate the characteristics and functionality of quantum batteries, numerous experiments have been carried out, using, for example, superconducting qubits~\cite{Hu_2022}, quantum dots~\cite{battery_dot}, and NMR~\cite{exp3,exp_M}.

Catalysts can be utilised for energy extraction from quantum batteries. In particular, initially, the battery and catalyst can be considered to be in a product state, then unitaries can be applied to the composite system, consisting of the battery and the catalyst, to extract energy from the battery~\cite{c_cat,rev_cat,coo_cat}. Ref.~\cite{rev_cat} investigated the role of uncorrelated state invariant catalysts in energy extraction from quantum batteries and showed that the maximum extractable energy from the battery in the presence of such catalysts is still bounded by its ergotropy. These are the catalysts which not only remain in the same state after the energy extraction as initially but also remain product with the battery at the end of the interaction. By losing the restriction of being a product at the end, in Ref.~\cite{coo_cat}, the authors showed that a state-invariant catalyst can transform any single-copy battery state to a corresponding completely passive state which has the same entropy as the initial battery state. {In this work we introduce a catalyst-like system that we term an ``energy-invariant assistant," which is an auxiliary system that helps to extract more energy from the battery than its ergotropy while keeping its own energy conserved after the energy extraction process. However, the final state of the assistant after energy extraction can be different from the initial. For energy extraction a global unitary is acted on it and the battery.}

To extract energy from the battery, we apply only those unitaries on the joint battery-assistant state which satisfy this property.
We show that corresponding to every battery there exists an energy-invariant assistant and a unitary using which one can always reach the ground state of the battery, proving its advantage over state-invariant catalysts, which can only transfer the battery state to a completely passive state that would be the same as the ground state only when the initial battery state is pure. {In addition, we discuss the dependency of complete energy extraction from the battery on the dimension of the assistant.} 

{If the global unitary used for energy extraction is energy-invariant, we cannot extract energy from a battery using an energy-invariant assistant. However, we show that this can be done in the presence of an additional system. We find that by using a proper additional system and an energy-invariant assistant, we can extract complete energy from the battery, where the composite dimension of the battery and the additional system is assumed to be equal to that of the assistant.}

Further, we provide a necessary and sufficient condition for a battery state to be unable to provide any energy with the assistance of an arbitrary but fixed energy-invariant assistants. We refer to such a pair of states of battery and assistant as inactive. One needs to keep in mind that inactivity of such states will depend on the corresponding assistant's state. Changing the assistant's state may make an inactive battery active. {In Ref.~\cite{exp}, authors experimentally demonstrate that both the stored and extractable energies of the quantum battery can be significantly enhanced by introducing a catalytic system.  They employed conventional quantum catalysts that return to their initial states after interaction. Since such catalysts clearly preserve energy of themselves, they can also function as energy-preserving assistants while facilitating energy transfer. Therefore, the experimental realization of catalytic processes of this kind is indeed feasible for the implementation of the type of assistant that we propose in our work.}

The rest of the paper is arranged as follows: In Sec.~\ref{s2}, we provide a brief description of the energy-invariant assistant. In Sec.~\ref{s3}, we discuss the complete extraction of energy in the presence of energy-invariant assistants. The  necessary and sufficient condition for an arbitrary but fixed energy-invariant assistants to be unable to squeeze any energy from a given battery is presented in Sec.~\ref{s4}. In Sec.~\ref{s5}, we show the dependency on Hilbert space dimension of the energy-invariant assistant in acquiring more energy than ergotropy. Finally, the concluding remarks are presented in Sec.~\ref{concl}.
\section{Energy-invariant assistants}\label{s2}

{A catalyst, as we know from chemical science, is a substance, the presence of which can enhance the speed of a reaction without changing its own properties. In quantum informational tasks, the concept of catalysis was first utilized by Jonathan and Plenio in Ref.~\cite{Sx1} by showing the presence of a catalyst can activate state transformations which were prohibited otherwise keeping the initial and final state of itself same (see also Ref.~\cite{sx2}). {The idea of catalysts has been extended to quantum thermodynamics and generic resource theories \cite{_berg_2014, Lostaglio_2015, Ng_2015, M_ller_2018, Boes_2020}. The presence of a catalyst has been shown to reduce the dissipation of entropy and heat for information erasure \cite{Henao_2023}. Catalysts can also be used to cool down systems in situations when it can not be cooled in the absence of catalysts due to unavailability of required resources \cite{Henao_2021}. In Ref. \cite{Lipka_Bartosik_2021}, authors have proved that in resource theories defined through majorization, any resourceful state, when available in multiple copies, can serve as a catalyst for any allowed transformation.}}
\\

\noindent \textbf{Energy-invariant assistant:} 
{In general, quantum catalysts retain their state before and after the quantum operation. In contrast, here we relax this requirement and only demand that the energy of the assistant remain unchanged following the operation. Systems satisfying this condition are referred to as energy-preserving assistants.}
Let us consider the initial states of a battery, $B$ and the {assistant}, $C$, to be $\rho_B$ and $\rho_C$, which act on the Hilbert spaces, $\mathcal{H}_B$ and $\mathcal{H}_C$, respectively. The joint Hilbert space of $B$ and $C$ is $\mathcal{H}_{BC}\equiv\mathcal{H}_B\otimes\mathcal{H}_C$. The local Hamiltonians are describing the energy of $B$ and $C$ are $H_B$ and $H_C$. We consider the initial joint state of $BC$ to be a product, i.e., $\rho_B \otimes \rho_C$. To extract energy from a quantum battery, we allow global unitary operation, $U$, on the whole system comprising $B$ and $C$ which may squeeze more energy $B$ than unitaries acting only on $B$.

In this section, we constrain the initial and final energies of composite system of $C$ to remain invariant. In particular, the relation 
\begin{equation}
    \tr[\tr_B[U (\rho_B\otimes \rho_C) U^{\dag}]H_C]=  \tr[\rho_CH_C],\nonumber
\end{equation}
needs to be satisfied $\rho_C$ to be an energy-invariant assistant. {
It is crucial to note that such operations cannot be achieved using energy-preserving unitaries. If the energies of both the assistant and the composite system comprising the battery and the energy-invariant assistant are conserved, the energy of the battery itself would necessarily remain unchanged after the action of the unitary operation. However, energy extraction using energy-preserving unitary becomes possible by introducing an additional system into the process.}

{Let the initial states of the battery and the additional system be $\rho_{B} = \sum_{ij} b_{ij} \ket{i}\bra{j}$ and $\rho_{A}=\ket{GS}\bra{GS}$ which act on the Hilbert spaces, $\mathcal{H}_B$ and $\mathcal{H}_A$, respectively. Here the additional system's state is consciously chosen to be in the ground state, $\ket{GS}$, of the battery's Hamiltonian, $H_B$. The composite system consisting of the battery and the additional part is $\rho_{BA}=\rho_B\otimes\rho_A$.
We consider the 
initial state of the energy-invariant assistant to be 
$\rho_{C} = \ket{GS}\bra{GS} \otimes \ket{\phi}\bra{\phi},$ 
acting on the Hilbert space $\mathcal{H}_C$, where $\ket{\phi} = \sum_i \sqrt{b_{ii}}\ket{i}$. 
Hence, the total initial state of the entire setup is 
$\rho_{BAC} = \rho_{BA} \otimes \rho_{C},$ 
which act on the Hilbert space $\mathcal{H}_B \otimes \mathcal{H}_A \otimes \mathcal{H}_C.$ 
We assume that the dimension of $\mathcal{H}_C$ is equal to that of $\mathcal{H}_A\otimes \mathcal{H}_B$. 
The total Hamiltonian of the battery and additional system is denoted by $H_{BA}$ and is given by 
}
{
\begin{equation}
    H_{BA}= H_B \otimes I_A + I_B \otimes H_A\text{, }
\end{equation}}
{where $H_B=\sum_{i=1}^{d_B}h_i\ket{i}\bra{i}$ and $H_A=h_{GS} \ket{GS}\bra{GS}$ are the local Hamiltonians of the battery and additional system, respectively. {Here, $h_{GS}$ is the ground state energy of the battery's Hamiltonian, $H_B$, i.e., $h_{GS}=\min_ih_i$.} On the other hand, the Hamiltonian of the assistant considered as}
{\begin{equation}\label{4xe}
        H_{C}= I_A \otimes H_B  + h_{GS} {|GS\rangle\langle GS|} \otimes I_B.\nonumber
\end{equation}}
{The total Hamiltonian of the battery, additional system, and assistant is given by
$H_{BAC}=H_{BA}\otimes I_C+I_{BA} \otimes H_C$.
From the expressions of the Hamiltonians, $H_{BA}$ and $H_C$, it is evident that the initial energies of the composite system of the battery and additional system and the energy-invariant assistant are the same, i.e.,}
{
\begin{equation}
    \Tr[\rho_{BA}H_{BA}]= \Tr[\rho_{C}H_{C}]=\sum_ib_{ii}h_i+h_{GS.}
\end{equation}\label{kornikar}
}

{Let us define a swap unitary $U^{swap}_{BA:C}$ acting on the Hilbert space $\mathcal{H}_B\otimes\mathcal{H}_A\otimes\mathcal{H_C}$, which swaps the states between the Hilbert spaces $\mathcal{H}_B\otimes\mathcal{H}_A$ and $\mathcal{H}_C$. The action of the swap gate $U^{swap}_{BA:C}$ on the entire system will provide the following final state}
{\begin{equation}
\rho'_{BAC}=U^{swap}_{BA:C}\left[\rho_{BA}\otimes\rho_{C}\right]U^{swap\dag}_{BA:C}=\ket{GS}\bra{GS}\otimes\ket{\phi}\bra{\phi}\otimes\rho_{BA}.
\end{equation}}
{As can be confirmed using the expression of the Hamiltonians, $H_{BA}$ and $H_C$, the energy of the composite system comprising the battery and the additional system is equal to that of the energy-invariant assistant. Furthermore, even after the action of the swap unitary, the energies of both the composite system (battery plus additional system) and the energy-invariant assistant remain unchanged individually. Consequently, the total energy of the battery, additional system, and energy-invariant assistant is conserved under the operation of $U^{\text{swap}}_{BA:C}$. Upon tracing out the additional system and the assistant, the reduced state of the battery becomes the ground state of the Hamiltonian, $H_B,$ i.e., $\rho'_B = \mathrm{Tr}_{AC}[\rho'_{BAC}] = |{GS}\rangle\langle{GS}|$. Therefore, we find that it is possible to extract complete energy from the quantum battery in the presence of an additional system by an energy-preserving global unitary and energy-invariant assistant where the total Hilbert space dimension of the composite system of the battery and additional system is equal to the Hilbert space dimension of the assistant.} 

{In the above-described process, the energy of the battery is transferred to the additional system. However, this can be done even without the presence of an energy-invariant assistant. In this work, we are interested in analyzing if an assistant by itself, due to its presence, can increase the amount of extractable energy from a battery. To ensure that the additional energy is being extracted from the battery itself and not from the assistant, we put the restriction of energy conservation on the assistant and allow energy non-preserving unitaries.}

    \begin{figure}
		\centering
	\includegraphics[scale=0.12]{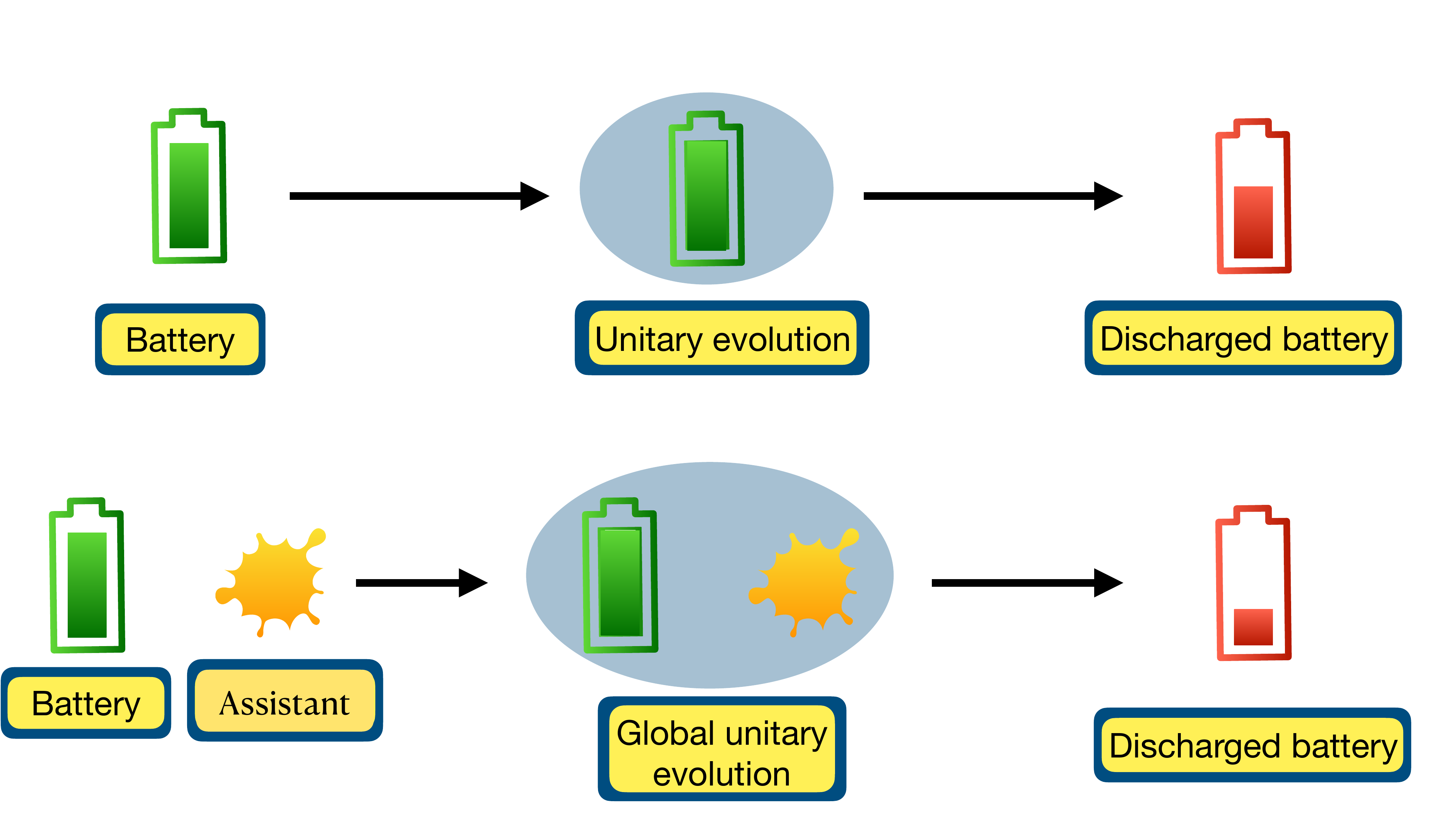}
		\caption{Schematic comparison between the energy extraction from a battery in the absence and presence of assistants. 
        In the absence of the assistant,  to extract energy from the battery, a unitary is applied directly on the battery. When the presence of an assistant is considered, the joint initial state of the battery and assistant is taken to be a product. A unitary is applied on the battery-assistant state to extract energy from the battery that preserves energy of the assistant. Though the energy of the assistant remains the same after the unitary evolution as its initial energy, implying no energy transfer from the assistant to the battery or vice versa, more energy can be extracted from the battery compared to the case where the assistant is not present.} 
		\label{fig:my_label}
	\end{figure}

The protocol of energy extraction from quantum batteries in the absence and presence of an energy-invariant assistant is illustrated in Fig. \ref{fig:my_label} using a schematic diagram. 

Similar to energy-invariant assistants, state-invariant catalysts can also be defined, which have been discussed in Ref. \cite{rev_cat,coo_cat} in the context of energy extraction from quantum batteries. It has been shown that in the presence of state-invariant catalysts, i.e., the catalysts that keep their state unaltered after their interaction with the battery to their initial state, any battery state can be transformed to its corresponding completely passive state, which has the same entropy as the initial battery state~\cite{coo_cat}. Hence, if the initial state of the battery is mixed, the final state of the battery, after the extraction of all possible energy, will remain mixed. For example, if in an extreme situation the initial state of the battery is maximally mixed, then even in the presence of a catalyst, the final state of the same will also be maximally mixed, implying zero extraction of energy. In this work, we first show that in the presence of an energy-invariant assistant, it's always possible to transform any battery's state (even if the initial state is maximally mixed) to its ground state, resulting in a non-zero amount of energy extraction, unless the initial state of the battery is in ground state. However, for such complete extraction of energy, we may need a particular state of the energy-invariant assistant depending on the initial state of the battery. For a fixed state of the battery, any arbitrary but specific state of energy-invariant assistant may not be suitable for squeezing energy from the battery. In the latter part of this paper, we find the necessary and sufficient condition for a battery-assistant state to be unable of providing any energy from the battery, keeping the energy of the assistant fixed.

From now on, throughout the paper we will use the notation $I_X$ to represent identity operator acting on the Hilbert space $\mathcal{H}_X$, where $X=B$, $C$, $BC$, would denote the individual Hilbert spaces of the battery and assistant or the joint Hilbert space describing the battery-assistant.
\section{Effectiveness of energy-invariant assistant
}\label{s3}

In this section, we focus on the efficiency of energy-invariant assistants in extracting energy from quantum batteries. Here we demonstrate that using an appropriate energy-invariant assistant and unitary, complete energy of a battery can be extracted, transforming the battery into its ground state. In particular, we confirm the following theorem:

\begin{theorem}
\label{th1}
Corresponding to every state, $\rho_B$, of battery, there exists a unitary, $U$, and energy-invariant assistant, prepared in a pure state $\ket{\phi_C}$, such that $\tr_C\left[U\left(\rho_B \otimes \ketbra{\phi_C}{\phi_C}\right) U^{\dag}\right]=\ketbra{\psi_g}{\psi_g}$ and the assistant being energy-invariant, satisfy $\tr\left[\tr_B\left[U (\rho_B\otimes \ketbra{\phi_C}{\phi_C}) U^{\dag}\right]H_C\right]=  \bra{\phi_C}H_C\ket{\phi_C}$. Here $H_C$ is the Hamiltonian of the assistant and $\ket{\psi_g}$ is the ground state of the battery.
\end{theorem}

\noindent \textit{Proof.}
We can fix the dimension of the Hilbert space describing the battery and the Hamiltonian of the battery as $d$ and  consider $H_B=\sum_{i=1}^d h_i \ket{\psi_i}\bra{\psi_i}$, where $\ket{\psi_i}$ is the eigenvector of $H_B$ with eigenvalue $h_i$ and the eigenvector $\ket{\psi_g}$ corresponding to $i=g$ is the ground state with energy $h_g$. 
We take the dimension of $\mathcal{H}_C$ to be the same as $\mathcal{H}_B$, which is $d$.
Let the Hamiltonian of the assistant be $H_C=\sum g_i\ket{\phi_i}\bra{\phi_i}$.

We confirm the above result in two steps. In the first step, we restrict to the initial states of the batteries that are incoherent in the energy basis and prove the stated theorem for them. Following this, in the second step, we generalize the proof for arbitrary states of the battery.\\

\noindent \textbf{Step 1.}
Let us consider the initial state, $\rho^{inco}_B=\sum_{i=1}^d \alpha_i\ket{\psi_i}\bra{\psi_i}$, of the battery, $B$, to be incoherent in the energy eigenbasis. Since $\{\alpha_i\}$ is the set of eigenvalues of $\rho^{inco}_B$, $\alpha_i\geq 0$ $\forall$ $i$.
Moreover, we consider the initial state of the assistant, $C$, to be a pure state of the form $\ket{\phi_C}=\sum_{i=1}^d \sqrt{\alpha_i}\ket{\phi_i}$.
Hence the energy of the initial state of the assistant in this case is given by
 \begin{eqnarray}
\bra{\phi_C}H\ket{\phi_C}& =&\sum_{i,j,k=1}^d\sqrt{\alpha_i}\bra{\phi_i}\left(g_k\ket{\phi_k}\bra{\phi_k}\right)\sqrt{\alpha_j}\ket{\phi_j}
=\sum_i \alpha_{i} g_{i}. \nonumber
 \end{eqnarray}
Here we have used the property that $\{\ket{\phi_i}\}$ forms an orthonormal basis, which implies $\braket{\phi_i}{\phi_j}=\delta_{ij}$, $\forall i,j$.

For extraction of energy, let us consider the global unitary, $U=(U_g\otimes I_C)(I_B \otimes U_k)U_{swap}=(U_g \otimes U_k)U_{swap}$. Here $U_g$ and $U_k$ are the local unitaries which act on, respectively, $\mathcal{H}_B$ and $\mathcal{H}_C$, in such a way that $U_g\ket{\phi_C}=\ket{\psi_g}$ and $U_k\ket{\psi_i}=\ket{\phi_i}$, $\forall i$ (we can define such a unitary because both $\{\ket{\psi_i}\}$ and $\{\ket{\phi_i}\}$ forms orthonormal basis). {One needs to keep in mind that since both $\mathcal{H}_B$ and $\mathcal{H}_C$ have the same dimension, the vectors $\ket{\phi_i}$, $\ket{\phi_C}$, $\ket{\psi_i}$ belong to both the Hilbert spaces, $\mathcal{H}_B$ and $\mathcal{H}_C$.} $U_{swap}$ is the global swap operator, which swaps the two systems, $B$ and $C$. The joint state of $B$ and $C$ obtained after applying the unitary is given by
 \begin{eqnarray}
&&U\left(\rho^{inco}_B \otimes \ket{\phi_C}\bra{\phi_C}\right) U^{\dag}\nonumber\\
&=&(U_g \otimes U_k)U_{swap}\left(\rho^{inco}_B \otimes \ket{\phi_C}\bra{\phi_C}\right) U_{swap}^{\dag}(U_g^{\dag} \otimes U_k^{\dag})\nonumber\\
&=&(U_g \otimes U_k)\left(\ket{\phi_C}\bra{\phi_C} \otimes  \rho^{inco}_B\right) \left(U^{\dag}_g \otimes U^{\dag}_k\right)\nonumber\\
&=&\left(  U_g\ket{\phi_C}\bra{\phi_C}U^{\dag}_g  \right)\otimes  \left(\sum \alpha_i U_k\ket{\psi_i}\bra{\psi_i}U^{\dag}_k\right) \nonumber\\
&=&\ket{\psi_g}\bra{\psi_g}  \otimes  \left(\sum \alpha_i \ket{\phi_i}\bra{\phi_i}\right). \nonumber
 \end{eqnarray}
The final state of $C$, after the action of the unitary, $U$, is $\widetilde{\rho}_C=\sum \alpha_i \ket{\phi_i}\bra{\phi_i}$. It is easy to check that the energy of $\widetilde{\rho}_C$ is $\tr[\widetilde{\rho}_C H_C]=\sum \alpha_i g_i$, which is the same as the energy of the initial state, $\rho_C$. Hence the assistant is energy invariant. On the other hand, the final state of the battery reaches the ground state of the battery's Hamiltonian, $H_B$. This proves that it is possible to extract full energy from the quantum battery when the initial state of the battery is in an incoherent state in the eigenbasis of the Hamiltonian of the battery.\\ 

\noindent \textbf{Step 2.}
In this step, we generalize our proof for arbitrary battery states, which may not be incoherent in the energy basis. Let the initial state of the battery be $\rho_B=\sum_i \alpha_i \ketbra{\xi_i}{\xi_i}$ and  $U_D$ be the unitary, which transforms $\ket{\xi_i}$ to $\ket{\psi_i}$, i.e., $U_D\ket{\xi_i}=\ket{\psi_i}$ $\forall i$. Hence $\rho^{inco}_B=U_D\rho_BU_D^{\dag}=\sum_i \alpha_i \ketbra{\psi_i}{\psi_i}$ is diagonal in the energy eigenbasis, $\{\ket{\psi_i}\}$. Here also, we consider the initial state of the assistant as $\ket{\phi_C}=\sum_{i=1}^d \sqrt{\alpha_i}\ket{\phi_i}$. Hence using $U_D$, we can diagonalize the state of the battery on the energy basis, and then, following the same method as before, we can extract the entire energy from $B$ keeping the energy of the assistant fixed. In particular, application of the unitary $U=(U_g\otimes U_k)U_{swap}U_D$ on $\rho_B\otimes \ketbra{\phi_C}{\phi_C}$ can transform $\rho_B$ to its ground state without changing the energy of the assistant.
We knew that using a state-invariant catalyst, we can always convert a battery's state to the completely passive state, which has the same entropy as the initial battery state~\cite{coo_cat}. Hence, unless the initial battery state is pure, it is not possible to transform it to its ground state with the help of a state-invariant catalyst. If the battery is in a pure state, the battery can be converted to its ground state by applying unitaries on itself even in the absence of any catalyst. Here we see, if we loosen the constraint on the catalyst to be state-invariant and only consider the systems whose only energy necessarily remains invariant after the interaction with the battery, with its initial energy, we can always reach the ground state of the battery.

\subsection{Example: Two level system}

In this subsection, we give an example of complete energy extraction from a battery using an energy-invariant assistant. We consider the Hamiltonians of the battery and assistant to be $\bar{H}_B=h_B \sigma_z$  and $\bar{H}_C=h_C \sigma_z$ respectively, where $\sigma_z$ is the Pauli matrix. We will express every matrix in the eigenbasis of $\sigma_z$. We choose the initial state of the battery to be
 $$
     \bar{\rho}_B=\left[
                \begin{array}{cc}
			\frac{1+k}{2} & 0 \\
			0 & \frac{1-k}{2}\\
			\end{array}
                \right]=\frac{1+k}{2}\ketbra{\bar{\psi}_e}{\bar{\psi}_e}+\frac{1-k}{2}\ketbra{\bar{\psi}_g}{\bar{\psi}_g}, 
$$ 
where $-1\leq k \leq 1$ and $\ket{\bar{\psi}_g}$ and $\ket{\bar{\psi}_e}$ are the ground and excited states of $B$.
The maximum extractable energy from this battery by applying unitary on the battery, i.e., the ergotropy of $\bar{\rho}_B$ is $2kh_B$ for $k>0$ and $0$ otherwise. We will show, in the presence of an energy-invariant assistant, $C$, $\bar{\rho}_B$ can be converted to $\bar{H}_B$'s ground state, extracting $(k+1)h_B$ energy. 
In this regard, we consider $\ket{\bar{\phi}_{C}}=\sqrt{\frac{1+k}{2}}\ket{\bar{\phi}_e}+\sqrt{\frac{1-k}{2}}\ket{\bar{\phi}_g}$ as the initial state of the assistant, where $\ket{\bar{\phi}_g}$ and $\ket{\bar{\phi}_e}$ are, respectively, the ground and excited states of $\bar{H}_C$. Let us take the unitary $\bar{U}=(\bar{U}_{g} \otimes I_C)\bar{U}_{swap}$, where
$$
\bar{U}_{swap}=\left[\begin{array}{cccc}
			1 & 0 &0 &0\\
			0 & 0 &1 & 0\\
            0 & 1 &0 & 0\\
            0 & 0 &0 & 1\\
			\end{array}\right] \text{ and }
            \bar{U}_{g}=\left[\begin{array}{cc}
			\sqrt{\frac{1-k}{2}} & -\sqrt{\frac{1+k}{2}}\\
			\sqrt{\frac{1+k}{2}} & \sqrt{\frac{1-k}{2}}\\
           \end{array}\right].
$$
The initial state of the composite system is $\bar{\rho}_B \otimes \ket{\bar{\phi}_{C}}\bra{\bar{\phi}_{C}}$. Applying $\bar{U}$, on the joint state of $B$ and $C$, we obtain the state $\ket{\bar{\psi}_g} \bra{\bar{\psi}_g} \otimes \rho_B$. 
Hence these specific choices of assistant's state, $\ket{\bar{\phi}_{C}}$, and unitary operator, $\bar{U}$, lead us to the extraction of the complete energy stored in the battery, i.e., $(k+1)h_B$.
\section{Necessary and sufficient condition of {inactivity} in the presence of energy-invariant assistants}\label{s4}

So far we have shown that there always exists an energy-invariant assistant and a unitary operator by which we can extract full energy from a quantum battery. However, in a practical scenario, one might not have access to the optimal state of the assistant, which provides complete energy extraction; instead, a particular state of the assistant may be available. 
This situation motivates us to find a necessary and sufficient condition for a battery to be incapable of providing any energy even in the presence of a given energy-invariant assistant prepared in an arbitrary but fixed state. {The passive states of a battery are states from which no energy can be extracted through any unitary operation acting solely on the quantum battery. Here, we do not apply a unitary operation on the battery alone. Instead, we apply it on the battery-assistant joint state. Owing to this distinction, we name the battery-assistant pair that cannot provide any energy by action of unitary on them as \textit{inactive}.}

We name such battery states as inactive. We would like to mention here that inactivity of a battery in presence of a assistant will depend on the state of the assistant. Altering the state of the assistant may convert the inactive battery to an active one.

\begin{theorem}
 \label{th2}
A battery, prepared in a state, $\sigma_B$, with Hamiltonian $H_{B}$, is inactive with respect to an energy-invariant assistant in state $\rho_C$, with Hamiltonian $H_C$, if and only if 
$C-xC{'}-\tilde{C}{''}\geq 0$,
where $C=H_{B_1} \otimes {I}_{C_1} \otimes \sigma_{B_2}^T \otimes \rho_{C_2}^{T}$, and 
$C{'}={I}_{B_1} \otimes H_{C_1} \otimes \sigma_{B_2}^T\otimes \rho_{C_2}^{T}$. Here the operators with subscript $B_m$ (${C_m}$) acts on the Hilbert space $\mathcal{H}_{B_m}$ ($\mathcal{H}_{C_m}$), that describes the battery (assistant), for $m=1$ and 2. The quantity, $x$ and $\tilde{C}{''}$, are defined as,
\begin{eqnarray}
   && x=\frac{\sum_i( \bra{i{i}}C\ket{\alpha {\beta}}-\bra{\beta {\alpha}}C\ket{i{i}})}{\sum_i( \bra{i{i}}C{'}\ket{\alpha {\beta}}-\bra{\beta {\alpha}}C{'}\ket{i{i}})}, \nonumber\\
   && \bra{\alpha {\beta}}\tilde{C{''}}\ket{\alpha' {\beta'}}=\delta_{\alpha \alpha'} \sum_i \bra{i {i}  } C-xC{'} \ket{\beta' {\beta}}\text{, }\nonumber
\end{eqnarray}
where $\{\ket{i}\}_{i}$ denotes the basis of $\mathcal{H}_{B_m}\otimes\mathcal{H}_{C_m}$ for both $m$=1 and 2. In denoting the joint basis element, $\ket{ij}$, we maintain the order of $\ket{i}$ and $\ket{j}$ such that the former (later) always acts on $\mathcal{H}_{B_1}\otimes\mathcal{H}_{C_1}$ ($\mathcal{H}_{B_2}\otimes\mathcal{H}_{C_2}$).
\end{theorem}
\noindent \textit{Proof.} A state, $\sigma_B$, of the battery would be inactive in the presence of an energy-invariant assistant, which is initially in a state, $\rho_C$, if for all $U$ which obeys the condition of energy-invariance of the assistant, $\rho_C$, i.e., 
\begin{equation}\label{5a}
    \tr\left[\tr_B\left[U (\sigma_B\otimes \rho_C) U^{\dag}\right]H_C\right]=  \tr[\rho_CH_C],
\end{equation} 
the following inequality is satisfied
\begin{eqnarray}
    \tr[\sigma_B H_B] &\leq& \tr\left[\tr_C\left[U (\sigma_B\otimes \rho_C) U^{\dag}\right]H_B\right] \nonumber \\
    &=& \tr\left[U(\sigma_B \otimes \rho_C)U^{\dag}(H_B \otimes {I}_C)\right].\label{myeq1}
\end{eqnarray}
The above relation infers the right-hand-side (RHS) of the relation is minimum for $U=I_{BC}$.

We will find the necessary condition of inactivity, by determining the maximum extractable energy (i.e., the minimum of RHS of inequality (\ref{myeq1})) from $\sigma_B$ using the assistant, $\rho_C$, and finding the necessary condition for the optimum to be reached at $U=I_{BC}$. Since we want to optimize over those $U$ which satisfy equation (\textcolor{red}{\ref{5a}}), we will use Lagrange's multipliers. Henceforth, in this section, we will consider the Choi Jamilkowski operator formalism to derive the conditions of inactivity.

RHS of inequality \ref{myeq1} can be 
written as 
\begin{equation}
    \small{\tr[U(\sigma_B \otimes \rho_C)U^{\dag}(H_B \otimes {I}_C)]]=\tr[C_{B_{1}C_{1}B_{2}C_{2}}E_{B_{1}C_{1}B_{2}C_{2}}], }\label{myeq9}
\end{equation}
 where $C_{B_{1}C_{1}B_{2}C_{2}}=H_{B_1} \otimes {I}_{C_1} \otimes \sigma_{B_2}^T \otimes \rho_{C_2}^T$ and $E_{B_{1}C_{1}B_{2}C_{2}}=U \otimes {I}_{B_2C_2}\left(\sum_{ij}\ket{jj}\bra{ii}\right)U^{\dag} \otimes {I}_{B_2C_2}$ is the Choi state corresponding to $U$.
The operators, $E_{B_{1}C_{1}B_{2}C_{2}}$ and $C_{B_{1}C_{1}B_{2}C_{2}}$, act on the Hilbert space, $\mathcal{H}_{B_1} \otimes \mathcal{H}_{C_1} \otimes \mathcal{H}_{B_2} \otimes \mathcal{H}_{C_2}$. The superscript,
${T}$, denotes the transpose of the corresponding operator. We will omit the subscripts in $C_{B_{1}C_{1}B_{2}C_{2}}$ and $E_{B_{1}C_{1}B_{2}C_{2}}$ further in our analysis for simplification of notation. 
Expressing $C$ as $C=\sum_{pqrs}c^{qp}_{rs}\ketbra{q}{p}\otimes\ketbra{r}{s}$, equation \ref{myeq9} 
can further be simplified as 
\begin{eqnarray}
&&\Tr{[CE]}\nonumber\\
&=&\sum_{ijpqrs}\tr\left[\left(U\ket{j}\bra{i}U^{\dag}\otimes \ket{j}\bra{i}\right)c^{qp}_{rs}\ketbra{q}{p}\otimes\ketbra{r}{s}\right]\nonumber\\\nonumber
&=&\sum_{ijpqrs}c^{qp}_{rs}\bra{p}U\ket{j}\bra{i}U^\dag\ket{q}\braket{s}{j}\braket{i}{r} \\\nonumber
&=&\sum_{pqij}c^{qp}_{ij}U_{pj}U_{iq}^\dag\\\nonumber
&=&\sum_{pqij}(R_{pj}+\iota M_{pj})(R_{qi}-\iota M_{qi})\bra{i} C \ket{j}_{qp},
\end{eqnarray}
where $R$ and $M$ are the real and imaginary parts of the unitary, $U$, and $U_{\alpha\beta}$, $R_{\alpha\beta}$, and $M_{\alpha\beta}$, denote the ${\alpha\beta}^{\text{th}}$ element of $U$, $R$, $M$, respectively. The notation, $\bra{i} C \ket{j}_{qp}$, denotes the complex number $c^{qp}_{ij}$ which is the $qp$th element of the operator $\bra{i} C \ket{j}=\sum_{pq}c^{qp}_{ij}\ketbra{q}{p}$ that acts on $\mathcal{H}_{B_1}\otimes\mathcal{H}_{C_1}$. $\iota$ is the imaginary number, $\sqrt{-1}$.

Our aim is to minimize the quantity, $\Tr{[CE]}$ (i.e., RHS of inequality \ref{myeq1}), over the set of real parameters $\{R_{\alpha\beta},M_{\alpha\beta}\}_{\alpha,\beta}$, subject to the two following sets of constraints: 
\begin{itemize}
    \item The first set of constraints arises from the definition of the unitary operators, i.e., $U^\dagger U=I_{B_1C_1}$. This relation can be expressed as  $\left(U^\dagger U\right)_{jk}=\delta_{jk}$ which imposes the following restrictions
    $$(R_{ij}-\iota M_{ij})(R_{ik}+\iota M_{ik})=\delta_{jk}$$
    on the parameters, $\{R_{\alpha\beta},M_{\alpha\beta}\}_{\alpha,\beta}$. 
    \item The second one arises from the assistant's energy-invariance restriction, which is given in equation~\ref{5a}. This relation can be rewritten as 
\begin{equation}   \sum_{ij}\Tr[U\ket{j}\bra{i}U^{\dag}\otimes \ket{j}\bra{i}C']=\sum_{ij}\Tr[\ket{j}\bra{i}\otimes \ket{j}\bra{i}C'], \label{myeq5}
\end{equation}
where  $C{'}\equiv C{'}_{B_{1}C_{1}B_{2}C_{2}}={I}_{B_1} \otimes H_{C_1} \otimes \sigma_{B_2}^T \otimes \rho_{C_2}^T$. Again by using the relation $U=R+\iota M$ and writing $C'$ as $C'=\sum_{pqrs}c'^{qp}_{rs}\ketbra{q}{p}\otimes\ketbra{r}{s}$ we can express the constraint in terms of the parameters $\{R_{\alpha\beta},M_{\alpha\beta}\}_{\alpha,\beta}$ as
\begin{eqnarray}\label{777}
    \sum_{pqij}(R_{pj}+\iota M_{pj})(R_{qi}-\iota M_{qi})\bra{i} C{'} \ket{j}_{qp}=\sum_{ij}\bra{i}C{'}\ket{j}_{ij},\nonumber
\end{eqnarray}
where $\bra{i}C{'}\ket{j}_{ij}=c'^{ij}_{ij}$ is the $ij$th element of the operator $\bra{i}C{'}\ket{j}$ that acts on $\mathcal{H}_{B_1}\otimes\mathcal{H}_{C_1}$.
\end{itemize}
Since we have now expressed the function to be optimized, as well as all the constraints, in terms of the same set of real parameters, $\{R_{\alpha\beta},M_{\alpha\beta}\}_{\alpha,\beta}$, we can now write the corresponding Lagrangian with undetermined multipliers. The Lagrangian of the constrained minimization problem is given by
\begin{eqnarray}
 \mathcal{L}
&=&\sum_{pqij}(R_{pj}+\iota M_{pj})(R_{qi}-\iota M_{qi})\bra{i} C{''} \ket{j}_{qp}\nonumber \\
&-&
\sum_{jk}\lambda_{jk}\left(\sum_{i}(R_{ij}-\iota M_{ij})(R_{ik}+\iota M_{ik})-\delta_{jk}\right)
+x\sum_{ij}\bra{i} C{'} \ket{j}_{ij},\nonumber
\end{eqnarray}
where $\lambda_{jk}$ and $x$ are the Lagrange's multipliers and $C{''}=C-xC{'}$.  The first order necessary conditions of obtaining a minimum of the Lagrangian, $\mathcal{L}$, are
$\frac{\partial \mathcal{L}}{\partial R_{\alpha,\beta}}=0$ and $\frac{\partial \mathcal{L}}{\partial M_{\alpha,\beta}}=0$. Taking partial derivative of $\mathcal{L}$ with respect to $R_{\alpha,\beta}$ and $M_{\alpha,\beta}$, $\forall \alpha, \beta$, we obtain
\begin{eqnarray}\label{dif_1}
\frac{\partial \mathcal{L}}{\partial R_{\alpha,\beta}}
=\sum_{iq} (R_{qi}-\iota M_{qi})\bra{i}C{''} \ket{\beta}_{q\alpha}-\sum_{k} \lambda_{\beta k}(R_{\alpha k}+\iota M_{\alpha k })\\\nonumber
+\sum_{pj} (R_{pj}+\iota M_{pj})\bra{\beta}C{''} \ket{j}_{\alpha p}-\sum_{j} \lambda_{j\beta }(R_{ \alpha j}-\iota M_{\alpha j}),\\
\label{dif_12}
\frac{\partial \mathcal{L}}{\iota\partial M_{\alpha,\beta}}=\sum_{iq} (R_{qi}-\iota M_{qi})\bra{i}C{''} \ket{\beta}_{q\alpha}+\sum_{k} \lambda_{\beta k}(R_{\alpha k}+\iota M_{\alpha k})\\\nonumber
-\sum_{pj} (R_{pj}+\iota M_{pj})\bra{\beta}C{''} \ket{j}_{\alpha p}-\sum_{j} \lambda_{j\beta }(R_{ \alpha j}-\iota M_{\alpha j}),\nonumber
\end{eqnarray}
respectively.
 Our aim is to find the condition under which the unitary being the identity operator, $I_{B_1C_1}$, reaches the minimum. Hence we make the partial derivatives, $\frac{\partial \mathcal{L}}{\partial R_{\alpha,\beta}}$ and $\frac{\partial \mathcal{L}}{\partial M_{\alpha,\beta}}$, zero for $R=I_{B_1C_1}$ and $M_{\alpha\beta}=\boldsymbol{0}$ and get 
\begin{eqnarray*}
    \frac{\partial \mathcal{L}}{\partial R_{\alpha,\beta}}\Bigg{|}_{R=I,M=\boldsymbol{0}}=\sum_{i}\bra{ii}C''\ket{\alpha\beta}&+&\sum_{j}\bra{\alpha\beta}C''\ket{jj}\\&-&\lambda_{\beta\alpha}-\lambda_{\alpha\beta}=0,\\
     -\iota\frac{\partial \mathcal{L}}{\partial M_{\alpha,\beta}}\Bigg{|}_{R=I,M=\boldsymbol{0}}=\sum_{i}\bra{ii}C''\ket{\alpha\beta}&-&\sum_{j}\bra{\alpha\beta}C''\ket{jj}\\&+&\lambda_{\beta\alpha}-\lambda_{\alpha\beta}=0.
\end{eqnarray*}
By adding and subtracting the above two equations we get  
\begin{eqnarray}
\lambda_{\alpha\beta}&=&\sum_i\bra{ii}C''\ket{\alpha\beta},\text{ and }\label{dif_2}\\ 
\lambda_{\alpha\beta}&=&\sum_j\bra{\beta\alpha}C''\ket{jj} \forall\text{ }\alpha\text{ and }\beta.\label{dif_3}
\end{eqnarray}
Since $C''=C-xC'$, from equations \ref{dif_2} and \ref{dif_3}, $x$, can be found to be 
\begin{equation}
    x=\frac{\sum_i\left( \bra{ii}C\ket{\alpha \beta}-\bra{\beta\alpha }C\ket{ii}\right)}{\sum_i( \bra{ii}C{'}\ket{\alpha \beta}-\bra{\beta\alpha }C{'}\ket{ii})}\text{, } \forall \alpha, \beta.
\end{equation}

The second order necessary condition for minima to occur at a particular point is that the Hessian operator, $\mathbbm{H}$, should be semipositive at that point. In this case, $\mathbbm{H}$ contains the second-order derivatives of the Lagrangian, $\mathcal{L}$, with respect to $R_{\alpha \beta}$ and $M_{\alpha \beta}$, i.e.,
$$
     \mathbbm{H}=\left[\begin{array}{cc}
			P_1 & Q\\
			Q^T & P_2\\
			\end{array}\right] ,
$$
where $\bra{\alpha\beta}P_1\ket{\alpha'\beta'}=\frac{\partial^{2}\mathcal{L}}{\partial R_{\alpha' \beta'} \partial R_{\alpha \beta}}$, $\bra{\alpha\beta}P_2\ket{\alpha'\beta'}=\frac{\partial^{2}\mathcal{L}}{\partial M_{\alpha' \beta'} \partial M_{\alpha \beta}}$, and $\bra{\alpha\beta}Q\ket{\alpha'\beta'}=\frac{\partial^{2}\mathcal{L}}{\partial M_{\alpha' \beta'} \partial R_{\alpha \beta}}$. Let us now introduce a state, $\ket{\Psi}=\frac{1}{\sqrt{2}}\left(\begin{array}{c}
    \ket{\psi}\\-i\ket{\psi}
\end{array}\right)$, where $\ket{\psi}$ is any arbitrary normalised state which belongs to the Hilbert space on which $P_1$, $P_2$, and $Q$ act (one needs to keep in mind here that $P_1$, $P_2$, and $Q$ have the same dimensions). From the condition, $\mathbb{H}\geq 0$, we have $\bra{\Psi}\mathbb{H}\ket{\Psi}\geq0$, which implies $\bra{\psi}P_1+P_2+\iota\left(Q^T-Q\right)\ket{\psi}\geq 0$. Since this relations holds true for any arbitrary $\ket{\psi}$ having the appropriate dimension, we find 
\begin{equation}
    P_1+P_2+\iota\left(Q^T-Q\right)\geq 0.\label{myeq2}
\end{equation}
Let us now focus on the individual operators, $P_1$, $P_2$, and $Q$.
By taking double derivatives of the Lagrangian, $\mathcal{L}$, and using the expressions of $\lambda_{\alpha\beta}$ as given in equations \ref{dif_2} and \ref{dif_3}, the elements of these operators can easily be found to be as
\begin{eqnarray*}
&&\bra{\alpha\beta}P_1\ket{\alpha'\beta'}=\bra{\alpha\beta}P_2\ket{\alpha'\beta'}\\&&\hspace{2cm}=\bra{\alpha \beta}C{''} \ket{ \alpha'\beta'}\nonumber+\bra{\alpha' \beta'}C{''} \ket{\alpha\beta}\\&&\hspace{2cm}-\delta_{\alpha\alpha'} \sum_i(\bra{ii}C''\ket{\beta \beta'} + \bra{\beta\beta'}C''\ket{ii}),\\
&&\bra{\alpha\beta}Q\ket{\alpha'\beta'}=\iota\bra{\alpha \beta}C{''} \ket{ \alpha'\beta'}-\iota\bra{\alpha' \beta'}C{''} \ket{\alpha\beta}\\&&\hspace{2cm}-\iota\delta_{\alpha\alpha'} \sum_i(\bra{ii}C''\ket{\beta \beta'}- \bra{\beta \beta'}C''\ket{ii}).
\end{eqnarray*}
At this moment, let us introduce three new operators, $\tilde{C}$, $\tilde{C}'$, and $\tilde{C}''$, defining them as $\bra{\alpha\beta}A\ket{\alpha'\beta'}=\delta_{\alpha\alpha'}\sum_i\bra{ii}A\ket{\beta\beta'}$, for $A=$ $\tilde{C}$, $\tilde{C}'$, and $\tilde{C}''$. From the relation, $C''=C-xC'$, we have
\begin{equation}
    \tilde{C}''=\tilde{C}-x\tilde{C}'. \label{myeq6}
\end{equation}
The matrices, $P_1$, $P_2$, and $Q$, can be expressed in terms of the newly defined operator, $\tilde{C}''$, as
\begin{eqnarray}
    P_1&=&P_2=C''+(C'')^T-\left( \tilde{C}''+ (\tilde{C}'')^T\right),\label{myeq3}\\
    Q&=&\iota\left(C''-(C'')^T\right)-\iota\left(\tilde{C}''-(\tilde{C}'')^T\right).\label{myeq4}
\end{eqnarray}
Substituting the expressions of $P_1$, $P_2$, and $Q$, from equations \ref{myeq3} and \ref{myeq4} to the condition \ref{myeq2}, we get 
\begin{equation}
    C''-\tilde{C}''\geq0 \hspace{0.8cm} \text{or} \hspace{0.8cm} C''\geq \tilde{C}'' .
\end{equation}
This is the final necessary condition for inactivity in the presence of a given energy-invariant assistant, as stated in the theorem.

Next, we want to prove that the above condition is also sufficient to certify the inactivity of a given battery state. In this regard, we assume $C''\geq\tilde{C}''$. Therefore, we can write
\begin{eqnarray}
\label{FF}
&&\sum_{ij}\tr[U\ket{j}\bra{i}U^{\dag}\otimes \ket{j}\bra{i}(C''-\tilde{C}'')]\geq 0,\nonumber\\
\text{or, }&&\sum_{ij}\tr[U\ket{j}\bra{i}U^{\dag}\otimes \ket{j}\bra{i}(C-\tilde{C})]\label{myeq7}\\
     &-&x\sum_{ij}\tr[U\ket{j}\bra{i}U^{\dag}\otimes \ket{j}\bra{i}(C^{'}-\tilde{C^{'}})] \ge 0,\nonumber
\end{eqnarray}
for any unitary, $U$ (here we have used equation \ref{myeq6}). However, we are allowed to use those unitaries which preserve the energy of the assistant, i.e., satisfy equation \ref{myeq5}. On the other hand, by definition, $\sum_{ij}\tr[\left(U\ket{j}\bra{i}U^{\dag}\otimes \ket{j}\bra{i}\right)\tilde{C}']=\sum_{ij}\left(\bra{i}U^\dagger\right)\bra{i}(\tilde{C}')(U\ket{j})\ket{j}=\sum_{ijp}\delta_{U\ket{i}U\ket{j}}\bra{pp}C'\ket{ij}=\sum_{ip}\bra{pp}C'\ket{ii}$. Here we have used the fact that if $U\ket{i}=U\ket{j}$ then $\ket{i}=\ket{j}$. Therefore, due to the restriction defined through equation \ref{myeq5}, the inequality given in \ref{myeq7} reduces to 
\begin{equation}
\sum_{ij}\tr[U\ket{j}\bra{i}U^{\dag}\otimes \ket{j}\bra{i}(C-\tilde{C})]\geq 0. \label{myeq8}
\end{equation}
Using the property $\sum_{ij}\tr[U\ket{j}\bra{i}U^{\dag}\otimes \ket{j}\bra{i}(\tilde{C})]=\sum_{ij}\left(\bra{i}U^\dagger\right)\bra{i}(\tilde{C})(U\ket{j})\ket{j}=\sum_{ip}\bra{pp}C\ket{ii}$ and recalling $E=U \otimes {I}\left(\sum_{ij}\ket{jj}\bra{ii}\right)U^{\dag} \otimes {I}$ we can express the inequality \ref{myeq8} as
\begin{eqnarray}   \Tr[CE]\geq\sum_{ip}\Tr[\ket{ii}\bra{pp}C]=\tr[\sigma_B H_B].
\end{eqnarray}
Hence the condition $C{''}-\tilde{C{''}}\geq 0$ reduces to equation \ref{myeq1} proving that $C{''}-\tilde{C{''}}\geq 0$ is also a sufficient condition of inactivity of a quantum battery in presence of a given energy-invariant assistant. This completes our proof.

\hfill $\blacksquare$
\\

\textbf{Remark.} Following the same logical flow, one can show, the relation $C''-\tilde{C}''\leq 0$ proves the corresponding battery assistant pair, say $\{\rho_B,\rho_C\}$, is most active, that is, there does not exist any $U$, which, in one hand, preserves the energy of the assistant and, in the other hand, increase the energy of the battery. In other words, any $U$, which maintains the energy invariancy of the assistant will either extract energy from the battery or keep it constant. 

\begin{figure}
		\centering
	\includegraphics[scale=0.62]{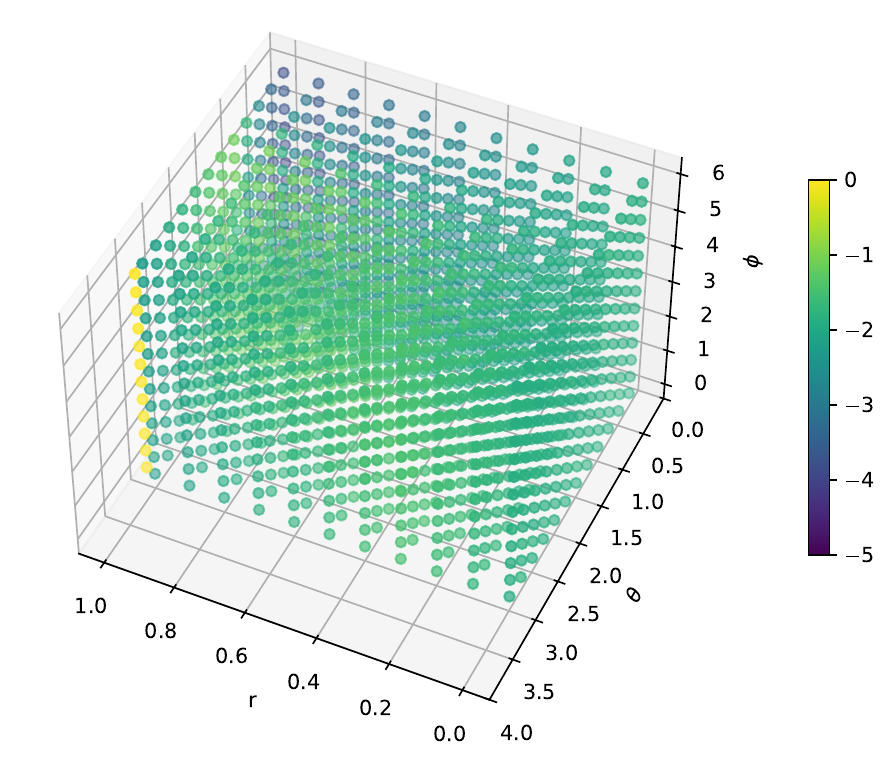}
		\caption{{Successful energy extraction from a qubit battery. The $X$, $Y$, and $Z$ axes denote the $r$, $\theta$, and $\phi$ parameters, respectively, that define the initial state, $\rho_B$, of the battery. Here the circular points denote the minimum eigenvalue of $C''-\tilde{C}''$, minimized over all qubit assistant states for given initial states, $\rho_B$. The negative minimum eigenvalues of $C''-\tilde{C}''$ indicates energy can be extracted from the corresponding battery state using a suitable assistant.}}  
		\label{fig:my_labelee}
	\end{figure}
{From theorem.~\ref{th2}, we can also claim numerically that there always exists a state-invariant assistant and a global unitary operation by which, from any battery state other than the ground state, energy can be extracted. To show this, we consider a qubit battery with initial state $\rho_B=\frac{(I+\vec{r}.\vec{\sigma})}{2}$. Here, $\vec{r}$ and $\vec{\sigma}$ are vectors with elements $\{r\sin(\theta)\cos(\phi),r\sin(\theta)\sin(\phi), r\cos(\theta)\}$ and $\{\sigma_x,\sigma_y,\sigma_z\}$, respectively. The range of $r$, $\theta$, and $\phi$ are given by $r\in[0,1], \theta \in [0,\pi]$, and $\phi\in[0,2\pi)$. At the same time $\sigma_x, \sigma_y$, and $\sigma_z$ are the Pauli matrices. The local Hamiltonians of the battery and assistant are chosen to be $\sigma_z$. According to the necessary and sufficient condition of inactivity, if for a given battery and assistant states $C''-\tilde{C}''\geq 0$, no energy can be extracted from the battery. Hence, if one of $C''-\tilde{C}''$ the eigenvalues turns out to be negative for a battery-assistant pair, then we can extract energy from the battery with the help of the assistant. Here we evaluate the minimum eigenvalue of the operator $C''-\tilde{C}''$ for a given battery state and further minimize that minimum eigenvalue over all possible single-qubit assistant states. In Fig.~\ref{fig:my_labelee}, we plot the minimum eigenvalue of the operator $C''-\tilde{C}''$ minimized over all possible assistant states with respect to the initial battery state parameters within the range $r\in[0,1]$, $\theta \in [0, 3]$, and $\phi\in [0, 6]$. We also plot the same considering the ground state of the battery for which $\theta=\pi$ and $r=1$. As the ground state is independent of the parameter $\phi$, there is more than one yellow point in the plot showing minimum eigenvalue $0$ all of which denoted. For other battery initial states, the minimum eigenvalues of the operator $C''-\tilde{C}''$, minimized over all assistant states, are clearly negative, as can be witnessed from the plot. It implies that the ground state is the only passive state. In other words, we can extract energy from any other state with the support of the auxiliary; hence, we can always reach the ground state of the battery considering appropriate energy-invariant assistants, which is previously indicated through Theorem 1.}

{\section{Assistants of different dimensions}\label{s5}}
{
Until now, we have considered the case where the energy-invariant assistant has a Hilbert space dimension equal to that of the battery. We now turn our attention to situations where the dimensions differ. To this end, we first examine the case in which the Hilbert space dimension of the assistant exceeds that of the battery. Let us consider the Hilbert space dimension of the battery and assistant, is $d_B$ and $d_C$, where $d_B<d_C$. The initial state of the battery is $\rho_B=\sum_{i,j}b_{ij}\ket{i}\bra{j}$. At the same time we choose the assistant's state as $\rho_C=\ket{\psi_C}\bra{\psi_C}$. Here, $\ket{\psi_C}$ is given by $\ket{\psi_C}=\sum_i\sqrt{b_{ii}}\ket{i}$, {where the summation in $i$ runs up to the Hilbert space dimension of the battery}. The local Hamiltonians of the assistant and battery are given by $H_B=\sum_ih^B_i\ket{i}\bra{i}$ and $H_C=\sum_j h^C_j\ket{j}\bra{j}$. The initial state of the composite system of battery and assistant is given by $\rho_{BC}=\rho_B\otimes\rho_C$ that acts on the Hilbert space $\mathcal{H}_B\otimes\mathcal{H}_C$. To extract energy from the quantum battery, we consider a global unitary acting on the Hilbert space of the composite system. The explicit form of the global unitary operator is given by
\begin{equation}
U = U_{GS}\otimes {I}\left(\begin{array}{cc}
U_{\text{Swap}}^{d_B^2 \times d_B^2} & X \\[6pt]
X' & X''
\end{array}\right)_{\!\!d_Bd_C \times d_Bd_C},
\end{equation}
where $X$, $X'$, and $X''$ are null matrices of appropriate dimensions and $U_{GS}$ is the unitary which can transform $\ket{\psi_C}$ to $\ket{GS}$. After the application of the unitary, $U$, the final state of the battery and the assistant is
\begin{eqnarray}
      \rho_{BC}'&=&U\rho_B\otimes \rho_CU^{\dag}=U_{GS}\ket{\psi_C}\bra{\psi_C}U^{\dag}_{GS}\otimes\left(\begin{array}{cc}
\rho_{B} & X \\[6pt]
X' & X''
\end{array}\right)\\\nonumber
&=&\ket{GS}\bra{GS}\otimes\left(\begin{array}{cc}
\rho_{B} & X \\[6pt]
X' & X''
\end{array}\right).
\end{eqnarray}
Consequently, it is possible to reach the ground state of the battery Hamiltonian even when the dimension of the energy-invariant assistant’s Hilbert space exceeds that of the battery’s Hilbert space, provided the initial state of the assistant is pure, that is, rank 1.
}

{On the other hand, when the Hilbert space dimension of the battery exceeds that of the assistant, i.e., $d_B > d_C$, two distinct situations can arise.
In one case, the rank of the battery, $r_B$, can be greater than the dimension of the assistant, $d_C$. In such a situation, the rank, $r_B$, can never be transferred to the assistant, because a state's rank can never be more than its dimension. As a result, the ground state of the battery cannot be reached, and complete energy extraction becomes impossible in this case using an energy-invariant assistant.
Conversely, when the rank of the battery is smaller than the assistant’s dimension, i.e., $r_B < d_C$, it becomes possible to transfer the battery’s entire rank to the assistant through a suitable unitary transformation. In this scenario, complete energy extraction can be achieved by subsequently applying a local unitary operation that brings the battery to the ground state of its Hamiltonian, provided the assistant is initially prepared in a pure state.}

{One can notice from the above discussion that the rank of the assistant plays a crucial role in complete energy extraction from the batteries. Therefore, if the assistant interacts with the environment, which brings decoherence to the assistant, the state of the assistant may become mixed. In such a case, the process of complete energy extraction using the noisy assistant state will fail. However, even in such scenarios, the activation of the assistant-battery pair, i.e., their ability to provide
energy, can be verified using the necessary and sufficient condition provided in Theorem 2.}

\section{Conclusion}
\label{concl}
We studied the role of an energy-invariant assistants in the extraction of energy from a quantum battery. In a previous work, considering state-invariant catalysts, the authors have shown that it is always possible to transform the state of the battery to a completely passive state, keeping the state of the catalyst unaltered at the end of its interaction with the battery. 

In this work, we loosened the restriction on the catalyst to be in the same final state as the initial, and constrained it to only have the same energy after its interaction with the battery as the initial. We named these systems as energy-invariant assistants. Considering such a scenario, we proved that for every battery, there exists an energy-invariant assistant in the same dimensional Hilbert space and a global unitary, the action of which on the battery-assistant state can extract the complete energy of the battery, leaving it in its ground state. Hence, in the presence of an energy-invariant assistant, not only a completely passive state but also the lowest energy state, that is, the ground state of the battery, is reachable.

{In addition, we have demonstrated that when the Hilbert space dimension of the energy-invariant assistant exceeds that of the battery, there always exists an energy-invariant assistant and a corresponding global unitary operation that enables complete extraction of energy from the quantum battery. Conversely, when the Hilbert space dimension of the assistant is smaller than that of the battery, two distinct scenarios arise. In the first, where the rank of the battery exceeds the assistant’s dimension, complete energy extraction becomes impossible. In the second, when the battery’s rank is smaller than the assistant’s dimension, full energy extraction from the battery remains achievable.}

However, there may exist situations where the optimal energy-invariant assistant is not available in the laboratory. In such a scenario, there exist inactive battery states, apart from the ground state of the battery, from which no energy can be extracted when a fixed state of energy-invariant assistants is available. This led us to find a necessary and sufficient  condition for an battery to be inactive, in the sense of unable to provide any energy, in the presence of an arbitrary but given energy-invariant assistant. 
\section*{Acknowledgment}
KS acknowledges support from the project MadQ-CM (Madrid Quantum de la Comunidad de Madrid) funded by the European Union (NextGenerationEU, PRTR-C17.I1) and by the Comunidad de Madrid (Programa de Acciones Complementarias). 

\bibliographystyle{iopart-num}

\bibliography{Battery}
\end{document}